\documentclass[graybox]{svmult}

\usepackage{type1cm}        %
\usepackage{makeidx}         %
\usepackage{graphicx}        %
\usepackage{multicol}        %
\usepackage[bottom]{footmisc}%

\usepackage{newtxtext}       %
\usepackage[varvw]{newtxmath}       %

\usepackage{graphicx} 
\usepackage{multicol} 
\usepackage{footmisc} 
\usepackage{color,xcolor}

\usepackage{hyperref}
\hypersetup{
    colorlinks=true,
    linkcolor=blue,
    filecolor=black,      
    urlcolor=blue,
    citecolor=blue,
    pdfpagemode=FullScreen,
    }

\makeindex             %

\begin{document}

\title*{Estimating Importation Risk of Covid-19 in Hurricane Evacuations: A Prediction Framework Applied to Hurricane Laura in Texas}
\titlerunning{Estimating Importation Risk of Covid-19 in Hurricane Evacuations} 

\author{Michelle Audirac$^\dagger$, Mauricio Tec$^\dagger$, Enrique García-Tejeda$^\ddagger$ and Spencer Fox$^\dagger$}
\authorrunning{Audirac et al.}
\institute{$^\dagger$The University of Texas at Austin \at
$^\ddagger$Centro de Investigación y Docencia Económicas}

\newcommand{\mauricio}[1]{{\color{brown}#1}}

\maketitle

\abstract*{
}

\abstract{
In August 2020 as Texas was coming down from a large summer COVID-19 surge, forecasts suggested that Hurricane Laura was tracking towards 6M residents along the East Texas coastline threatening to spread COVID-19 across the state and cause pandemic resurgences. To assist local authorities facing the dual threat, we integrated survey expectations of coastal residents and observed hurricane evacuation rates in a statistical framework that combined with local pandemic conditions predicts how COVID-19 would spread in response to a hurricane. For Hurricane Laura, we estimate that 499,500 [90\% Credible Interval (CI): 347,500, 624,000] people evacuated the Texan counties, that no single county accumulated more than 2.5\% of hurricane evacuees, and that there were 2,900 [90\% CI: 1,700, 5,800] exportations of Covid-19 across the state. In general, reception estimates were concentrated in regions with higher population densities. Nonetheless, higher importation risk is expected in small Districts, with a maximum number of importations of 10 per 10,000 residents in our case study. Overall, we present a flexible and transferable framework that captures spatial heterogeneity and incorporates geographic components for predicting population movement in %
the wake of a natural disaster. As hurricanes continue to increase in both frequency and strength, our framework can be deployed in response to anticipated hurricane paths to guide disaster preparedness and planning.
}

\section{Introduction}
\label{sec:intro}

Identifying which populations travel and how they spread in an evacuation is essential to assess the dual threat of a hurricane and an ongoing epidemic, yet anticipating hurricane evacuation is challenging. The relation between spatial patterns of storm surges, the geographic distribution of potential destinations, and the decisions that a household makes when threatened by an approaching hurricane is a highly complex and dynamic process \cite{baker}. Our aim is to offer the ability to anticipate individuals’ movement in an  evacuation before a hurricane hits land in the presence of an already prevalent regional epidemic, and quantify the importation and exportation of cases in both the origin and destination counties.

Geospatial Modeling of Storm Surges and other GIS tools have become an important tool in assessing the risk of hurricanes \cite{ferreira}. For this work, we examine existing models that combine geospatial features, storm surge risk, and statistical methods to integrate a modeling framework that, along with prevalence numbers of an infectious disease, can estimate importation risk during a hurricane evacuation. %
We apply our framework to the evacuation in the Texas region in response to the approach of Hurricane Laura during the Covid-19 pandemic. 
While the results presented here pertain to these specific geographic features, demographic characteristics, hurricane, and epidemic settings, the framework itself is flexible for application in future hurricane events and other geographical locations undergoing an epidemic.

The remainder of this study is organized as follows. First, we provide a general description of the events surrounding Hurricane Laura---including the prevalence of Covid-19 cases in Texas at the time of the approach of the hurricane and an account of which counties issued mandatory or voluntary evacuation orders. Next, we discuss the selection of models that integrate our proposed framework, followed by the results of their implementation in our case study. To highlight the applicability of the framework to other scenarios, a counterfactual scenario is explored where the forecast track of Laura is assumed to force the mandatory evacuation of an additional group of counties, corresponding to those regions where hurricane warnings were issued for Rita in 2005. %
Finally, we discuss the implications and limitations of the proposed framework.

\section{Case Scenario and Study Area}
\label{sec:case_study}

\begin{figure}
\centering
\begin{minipage}{.5\textwidth}
  \centering
  \includegraphics[width=.9\linewidth]{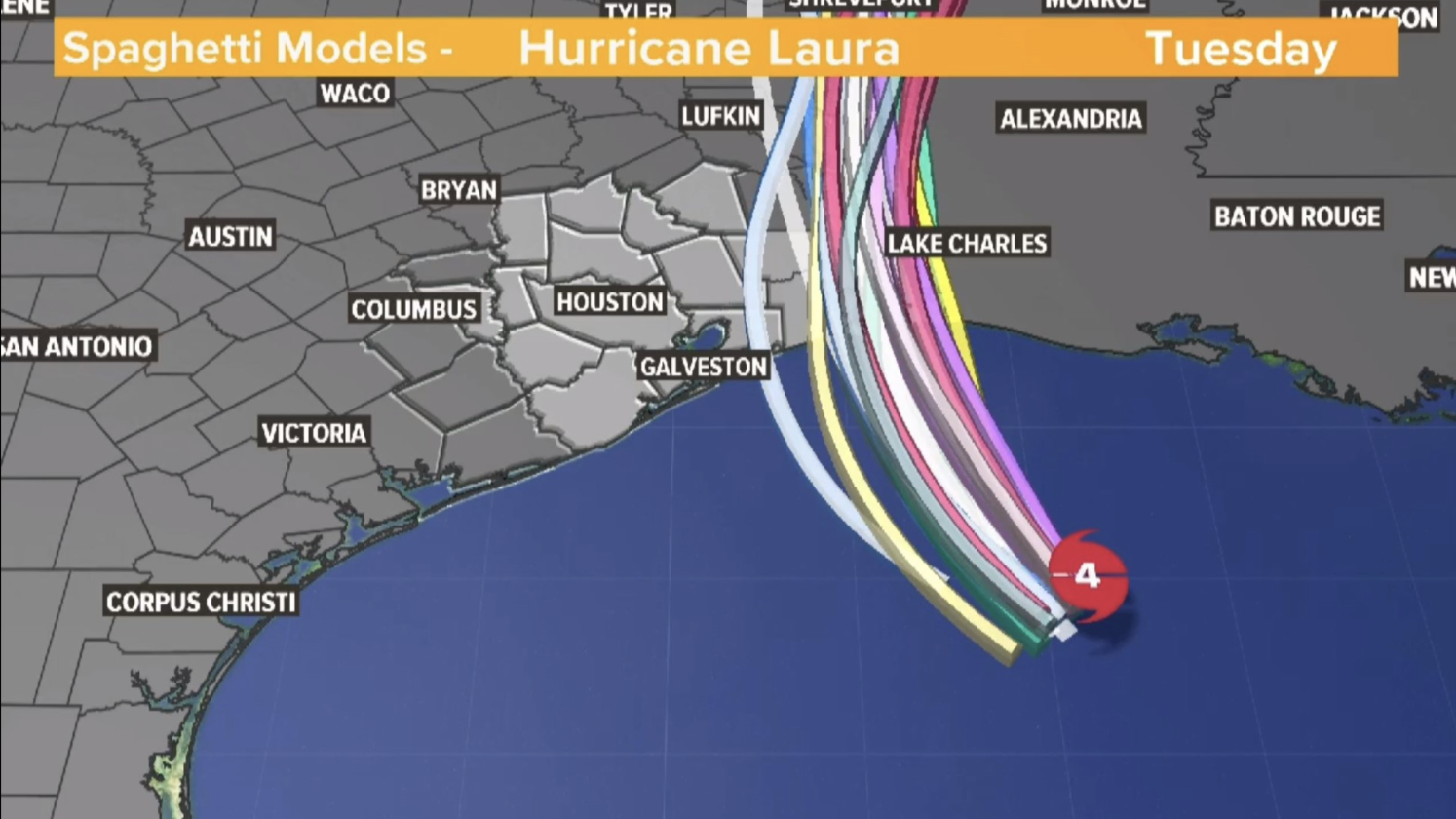}
\end{minipage}%
\begin{minipage}{.5\textwidth}
  \centering
  \includegraphics[width=.9\linewidth]{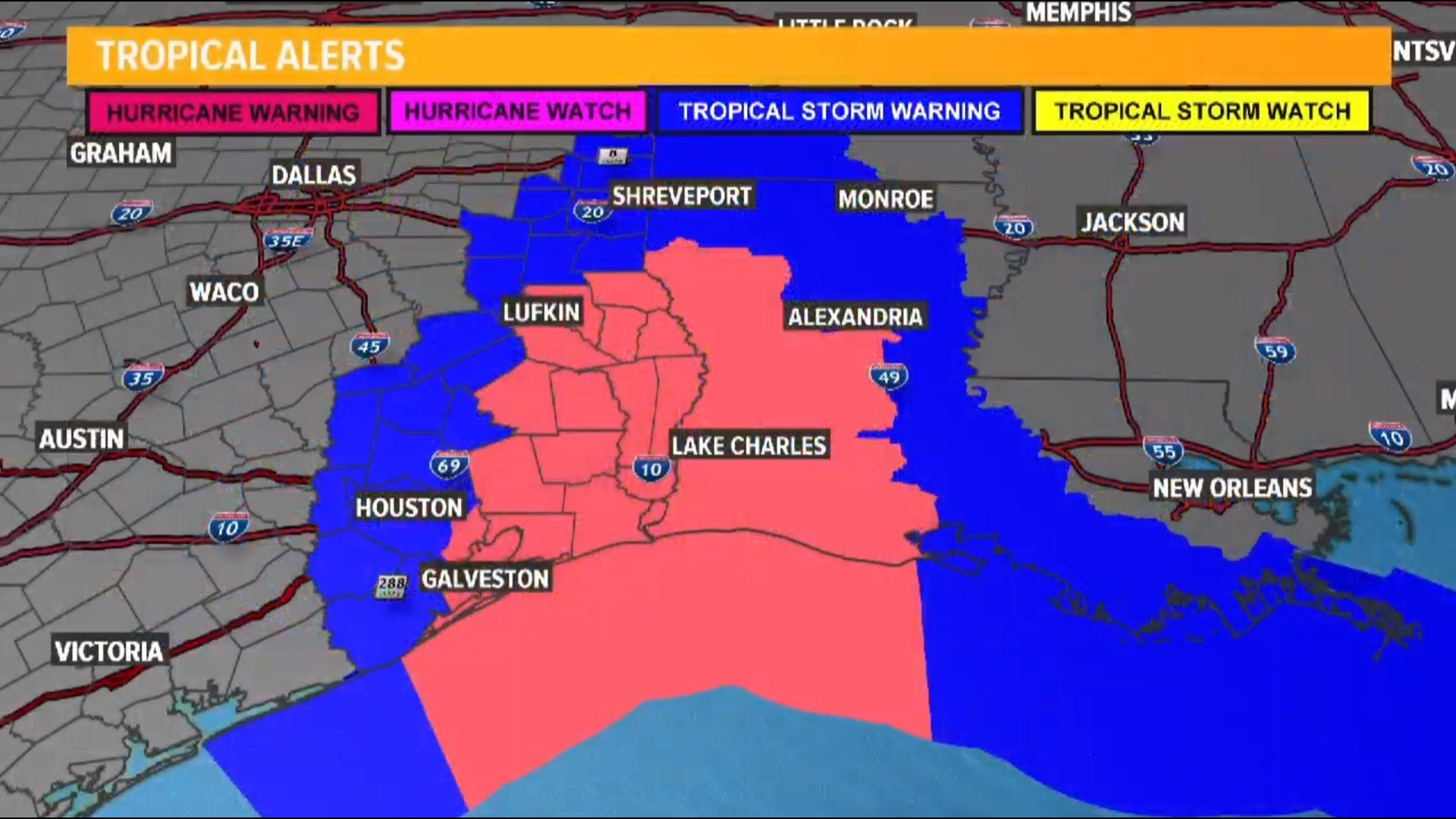}
\end{minipage}
\caption{The projected trajectory of Laura the day before the predicted landfall and the layout of counties that issued a hurricane warning (KHOU News \cite{khou}).}
\label{fig:khou}
\end{figure}

In August 2020, Hurricane Laura threatened 6M residents of the eastern Texas coastline amid the COVID-19 pandemic. To identify counties threatened by the forecast track of Laura, we used news coverage and State Situation Reports. A Hurricane Warning is declared in a region when the onset of hurricane conditions is likely within 24 hours. Figure \ref{fig:khou} shows the projected trajectory of Laura the day before the predicted landfall and the layout of counties that issued a hurricane warning.Within the highlighted region, 11 Texan counties issued mandatory evacuation orders, and 3 issued voluntary orders.

\begin{figure}
    \centering
    \includegraphics[width=.9\linewidth]{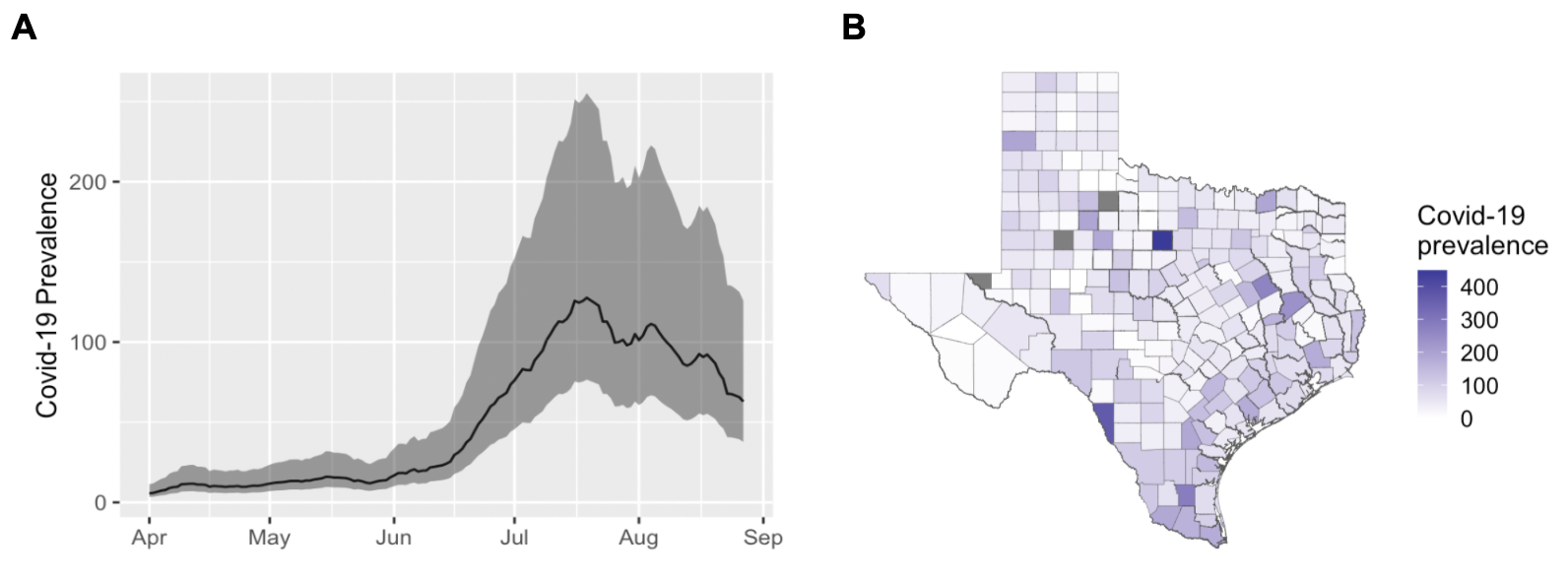}
    \caption{Estimated Covid-19 prevalence per 10,000 people in Texas: A) since April 2020, around the first identified Covid-19 case in the US. B) spatial distribution on the week prior to Hurricane Laura’s landfall on August 2020.}
    \label{fig:prevalence}
\end{figure}

During the days preceding Hurricane Laura’s landfall and all through August, the second surge of cases in Texas appeared to be mostly under control. Covid-19 cases in both the coastal counties and inland Texas had a downwards trend.  However, since the reported number of cases do not necessarily reflect the absolute number of people infected, we used the method by Fox et al. \cite{fox} to estimate a lower, median and upper bound of disease prevalence. The incidence estimation assumes that confirmed COVID-19 case counts represent somewhere between one third and one tenth of all infections. Figure \ref{fig:prevalence}a shows the Covid-19 prevalence per 10,000 residents in Texas from the beginning of the pandemic to the end of September, and panel \ref{fig:prevalence}b displays its spatial distribution estimated for the week of Aug 20-27. Overall, the peak of estimated cases is around 100 infections per 10,000 in mid August, with a substantial upper bound of up to 200 infections. The figure also shows that at that time, counties with higher prevalence were present in some inland regions, and not coastal areas.

\section{Background}
\label{sec:background}

In this section we examine existing models in the literature and specifically consider those that incorporate spatial variables and capture geographic and demographic heterogeneity. Because the aim of the proposed framework is to provide an agile response during future dual disasters, emphasis is put on the selected methods' ability to be applied over regions with varying geographic attributes, population distributions, as well as different hurricane strengths and levels of disease prevalence.

\subsection{Evacuation Rates}
\label{subsec:evac_backgroud}

Methods that focus on estimating %
evacuation rates fall mainly within two categories \cite{wilmot}: (1) household models and (2) participation rates approaches. %
In the case of household models, the primary purpose is identifying associations between household attributes and the decision of whether to evacuate. %
These attributes include demographic features, risk perception, and hurricane characteristics. %
These studies provide valuable knowledge about human decision-making under a hurricane emergency situation. %
However, most of the household models do not apply to an approaching hurricane event, nor do they take into account the geographic characteristics of the threatened regions and varying levels of surge or storm risk.

In the participation rates approaches, geographic areas are assigned estimates of evacuation rates based on key features such as hurricane strength and surge probability. The Federal Emergency Management Agency (FEMA) owns a comprehensive proprietary solution called Evacuation Traffic Information System (ETIS) \cite{etis} that, along hurricane traffic simulations, computes evacuation rates.
The applicability of the participation rate approaches make them more  suitable than household models for a framework whose emphasis is on preparation for specific storm. Nonetheless, access to previously trained participation rates methods is not publicly available.

\subsection{Destinations}
\label{subsec:dest_backgroud}

Two decisions lead to choosing a particular evacuation target location: (1) the accommodation type choice; and (2) the destination choice. An accommodation type is a sheltering option. Common alternatives include houses of friends and relatives, hotels, and public shelters \cite{mesa}. A destination choice, in constrast, is concerned with the flow of evacuees among predefined evacuation regions, or with the zone distribution of their accomodation. %

\runinhead{Accomodation type choice} Mesa-Arango et al. \cite{mesa} highlight the lack of evidence and prior rigorous statistical models explaining accommodation type demand. And, as far as we are aware, their work is the only one providing a statistical approach to model the accomodation type choice. Instead, the proportion of evacuees traveling to each accommodation type is typically a fixed proportion informed from a specific past hurricane.%

In terms of applicability, Mesa-Arango’s model \cite{mesa} is not appropriate for prediction in emergency situations because several of its input variables, such as previous experience with hurricanes, or being required to work during the evacuation, depend on targeted survey responses.

\runinhead{Destination choice} Among studies modeling the destination choice, three types of models can be discerned: the gravity model \cite{modali}, intervening opportunity model \cite{wilmot}, and multinomial logit model \cite{cheng}. These studies estimate zonal distribution of hurricane evacuation trips using a telephone survey conducted in South Carolina following Hurricane Floyd \cite{dow}, thereby obtaining two origin-destination (OD) matrices: one for evacuees going to friends and family; and the other for those staying in hotels. Both the gravity and the intervening opportunity models learn the influence of some measure of travel impedance, such as a function of travel time and distance, on the distribution of origin-destination trips. In the multinomial logit model, the probability of choosing a destination from a given location is estimated by incorporating demographic characteristics of destinations. 

Our approach to destination choice modeling follows that of Cheng et al. \cite{cheng}, by applying one of two multinomial logit regression models depending on the accomodation type (family/friends, hotels). Each incorporate demographics and travel impedance covariates easily obtainable from publicly available data.

\section{Modeling Framework}
\label{sec:model_frame}

We package selected methods into modules and build a framework that consists of two modules %
aimed at the following objectives:

\begin{enumerate}
    \item The estimation of evacuation rates per coastal county.
    \item The prediction of destination choices to obtain expected receptions in inland counties.
\end{enumerate}

Local prevalence estimates of COVID-19 are combined with evacuations and reception numbers from the framework's modules to determine which counties are likely to face the highest importation risk.

\subsection{Evacuation Rates Module}
\label{subsec:evac_module}

The first module of the framework consists of a participation rates approach that determines evacuation rates based on hurricane intensities and evacuation zone categories---which are geographic areas distinguished by their vulnerability to hurricanes of different strengths. Even though there is no convention on how jurisdictions determine hurricane evacuation zones, many jurisdictions use National Oceanic and Atmospheric Administration (NOAA)'s SLOSH surge model \cite{slosh} (see Figure \ref{fig:beta}a) as we do for our analysis. Below we briefly provide an overview of the approach. The reader should refer to Section \ref{appendix:beta} in the Appendix for additional details.

The proposed statistical model is a weighted generalized linear regression, informed by both evacuation expectations of coastal residents and observed compliance of evacuation orders in the region. The dependent variable considered is the probability of evacuation rate, modeled as a conditional Beta distribution, while the predictors are the risk zones and the hurricane categories.

To obtain data for estimating the parameters in the model, we compiled a database from multiple previous studies on Hurricane evacuations in the Gulf of Mexico area. These studies are \cite{data1, data2, data3, data4, data5} and \cite{data6}. The resulting database
---available with the code repository accompanying this paper at {\small\url{https://github.com/audiracmichelle/hurricane_covid.git}} ---contains $45$ entries.

Observations in the dataset fall mainly within two categories: first, 30 instances where the reported rate is the actual observed evacuation rate from a retrospective study; second, 15 cases where the recorded rate is based on reported intentions to evacuate based on survey responses. While both categories are included in the model to increase the learning data available, the responses based on intentions are given less weight in the regression than those from actual observations.

 The predicted conditional evacuation rates estimated from the model are shown in Figure \ref{fig:beta}b. The results confirm and quantify our intuition that the highest rate occurs in those zones that are most vulnerable and these rates are further inflated by the hurricane intensities.

Regions under mandatory evacuation orders are assigned an estimate of  total number of evacuees by multiplying the population in each evacuation zone by the evacuation rate corresponding to their surge risk and hurricane category. Regions under voluntary evacuation orders, are classified as risk zones for Hurricane category 0. The number of exportations is roughly the number of evacuees times the disease prevalence in a given area.

\begin{figure}[tbh]
    \centering
    \includegraphics[width=.8\linewidth]{figures/beta.png}
    \caption{A) Coastal zones are assigned a risk level using NOAA's SLOSH model \cite{slosh}. A risk zone delineates the near worst–case scenario of flooding for each hurricane category. Hurricanes categories are in order of increasing intensity. A risk area 3 has the potential of flooding with hurricanes of category 3, 4 and 5. B) Regression results showing an inverse relationship between risk zone and evacuation rates and direct relationship between hurricane category and evacuation rates.}
    \label{fig:beta}
\end{figure}

\subsection{Destinations Module}
\label{subsec:destin_module}

For this module we adopt the model of Cheng et al. \cite{cheng} whose  multinomial logit regression
uses a set of attributes that influence the preference for a destination alternative according to the accommodation type
as listed in Table \ref{tab:logit}. 

The values of the covariates for our case study are obtained from three main sources: (1) geographic features from the R package \texttt{geosphere} \cite{geosphere}; (2) demographic characteristics from the American Community Survey \cite{acs}; evacuation routes in Texas from the Homeland Infrastructure Foundation-Level Data API \cite{arcgis}, and hurricane forecast tracks from the National Hurricane Center GIS products \cite{NHC}. The values for the regression coefficients of these predictors are taken directly from the estimates in \cite{cheng}.

\begin{table}[tbh]
\caption{Variables used in the Origin-Destination Model.
}
\label{tab:logit}       %
\begin{tabular}{p{5.5cm}p{5.5cm}}
\hline\noalign{\smallskip}
\textbf{Friends/Relatives:} & \textbf{Hotel:} \\
\noalign{\smallskip}\svhline\noalign{\smallskip}
Miles from origin to destination & Miles from origin to destination \\
Destination population & Destination number of hotels\\
Hurricane threatened area (binary variable) & Hurricane threatened area (binary variable)\\
Metropolitan (MSA) indicator & Interstate highway indicator\\
Percentage white & Percentage white\\
\noalign{\smallskip}\hline\noalign{\smallskip}
\end{tabular}
\end{table}

Following \cite{data1}, we established fixed weights for the preference for accommodation type, with 60\% of evacuees going to family and friends and the remaining 40\% staying in hotels. Kang et al. \cite{data1} derived these proportions based on survey responses.
The model output consists of two origin-destination (OD) matrices, one for each accommodation type. Using the exportation numbers from the evacuation rates module and combining them with the estimated OD matrices gives the expected number of total importations for each evacuation destination choice. Section \ref{appendix:od} in the Appendix presents additional details on the statistical model.

\section{Results}
\label{sec:results}

Our focus is the preparation for specific approaching hurricanes, therefore the proposed framework is applied to the scenario where Hurricane Laura is forecasted to make landfall within the next 24 hours. In this timeframe, Laura was expected to reach the shore with a category-4 strength with a total of 6,016,750 residing in the counties under a hurricane warning. A total of 463,473 and 707,224 residents living in the surge risk zones are assumed to be under mandatory and voluntary orders to evacuate.

We obtain estimates of evacuation rates for census block groups (CBG) in the region as described in the Methodology section. The median predicted evacuation rate of CBG’s in risk zones with mandatory evacuation orders is 85\%, with 90\% the distribution of CBG predictions ranging from 69\% to 95\%. In contrast, CBG’s risk zones with voluntary orders had a median predicted evacuation rate of 6\%, and ranged between 3\% and 19\%. After aggregating the CBG results in a county we find that Orange County has the highest estimated evacuation rate, which is predicted to be 80\%. 

\begin{figure}
    \centering
    \includegraphics[width=.8\linewidth]{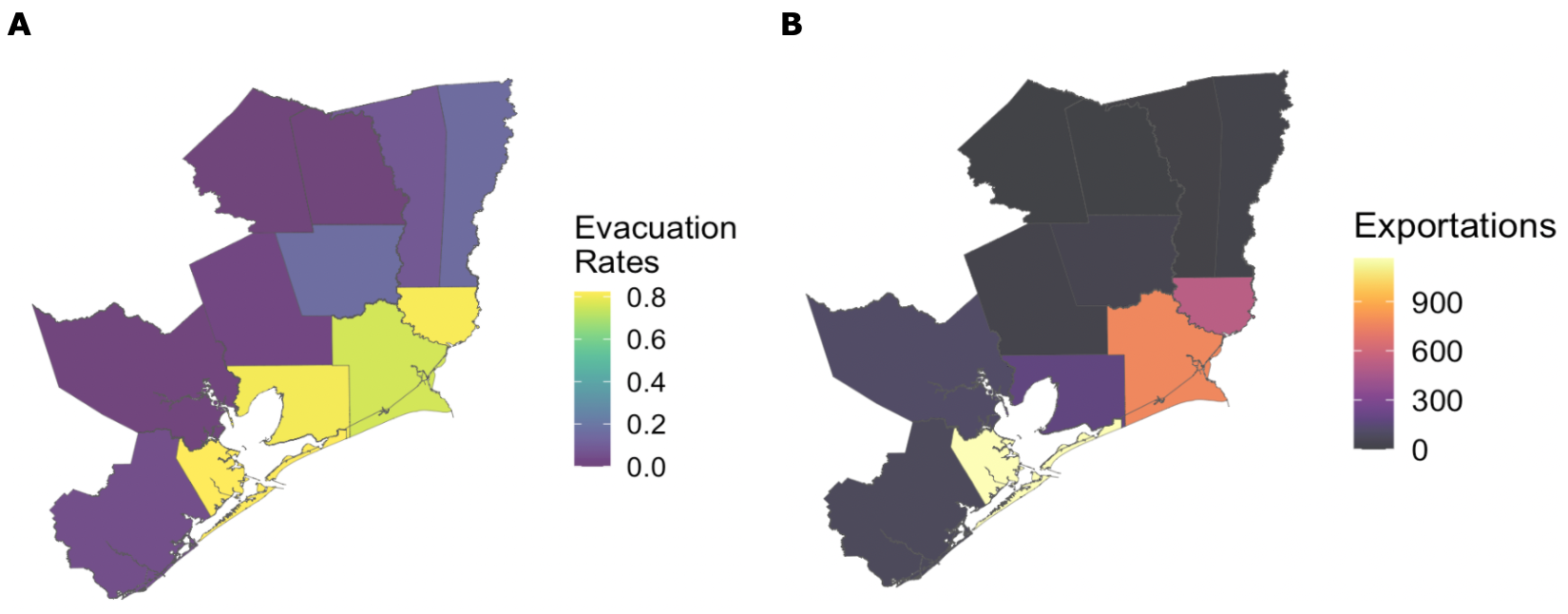}
    \caption{A) County-level evacuation rate prediction. Evacuation rates for counties were computed as the weighted averages of the census tract rates, where the weight for each tract was the percentage of the region population it contains.  B) Estimated number of total exported infections by county. Disease prevalence is estimated assuming 20\% infection detection rate. Evacuation rates and prevalence jointly determine the number of total exportations.}
    \label{fig:exportations}
\end{figure}

Multiplying the population in each county by the appropriate evacuation rate and disease prevalence provides an estimate of the total number of exportations. We estimate that 499,500 [90\% CI: 347,500, 624,000] people evacuated the Texan counties that issued an evacuation order. With prevalence in the region revolving around 66 per 10,000 people, the total number of predicted Covid infection exportations is 2,900 [90\% CI: 1,700, 5,800]. Figure \ref{fig:exportations} contains the spatial distribution of the county-level evacuation rates and exportation's estimates.

For each accommodation type, assuming 60\% of evacuees go to hotels and 40\% to family/relatives, we obtain a mapping of probabilities for all origin and destinations counties using the multinomial logit model described in the Methodology section. These probabilities along with the predictions for the number of evacuees captured by the evacuation rates module produce estimates of receptions for the destination counties.

Results show that travels disperse outside the threatened region; thus, no single county accumulates more than 2.5\% of the evacuees. When outcomes are aggregated in District level, the number of receptions (and receptions relative to the population of the district) in major districts include: San Antonio 44,800 (1.7\%), Austin 45,500 (1.9\%), Dallas 32,500 (0.65\%), and Fort Worth 29,000 (1.1\%). Relative to its population, reception of evacuees for counties in Yoakum District (Austin, Calhoun, Colorado, DeWitt, Fayette, Gonzales, Jackson, Lavaca, Matagorda, Victoria and Wharton) is of 17\%, which is the highest expected value in the case scenario.

\begin{figure}
\centering
\includegraphics[width=1\linewidth]{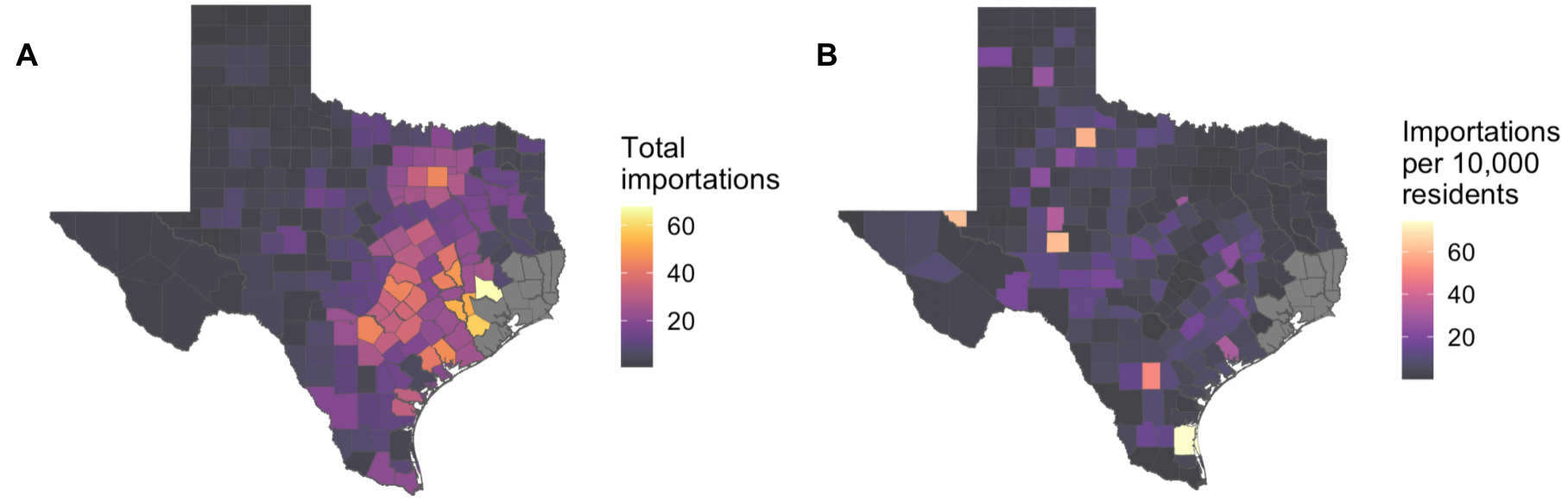}
\caption{Importation risk distribution: A) Estimated imported infections per destination county. B) Importations per 10,000 residents. In general evacuees were likely to move to regions with higher population densities. When aggregated in Districts, importations represent at most 10 infected people per 10,000 residents not exceeding 2 per 10,000 residents in the most populous Districts.}
\label{fig:destinations}       %
\end{figure}

Analogous results of imported infections show that county importations range between 3 and 18 (IQR); and no county imports more than 68 infections in total. Importation estimates (and importation per 10,000 residents) aggregated in Districts are as follows: San Antonio 257 (0.98), Austin 261 (1.11), Dallas 189 (0.35), and Fort Worth 169 (0.64). In Yoakum importations are estimated to be 333, corresponding to 10 importations per 10,000 residents. The dispersion of importations in the destination counties appears in Figure \ref{fig:destinations}.

\begin{figure}
    \centering
    \includegraphics[width=.7\linewidth]{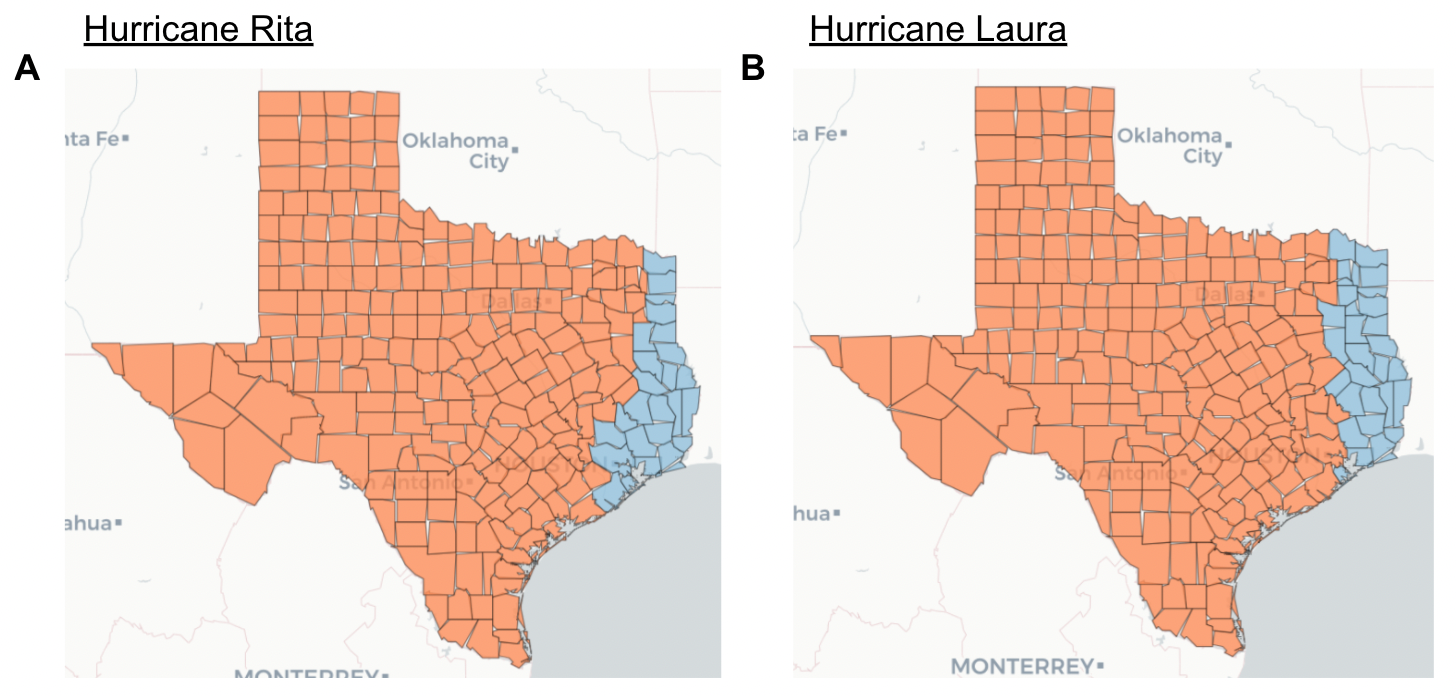}
    \caption{Counties in the path of the forecast tracks of both Hurricane Laura (2020) and Hurricane Rita (2005), representing the case study and counterfactual scenarios of our study.}
    \label{fig:rita}
\end{figure}

To validate and assess the transferability of the framework, we simulate a counterfactual scenario with conditions similar to those of Hurricane Rita. This hurricane, of category 5, had a feared forecast track pointing directly towards the highly populated area of Houston/Galveston.
Counties in the forecast tracks of both hurricanes appear in Figure \ref{fig:rita}. To fix the set of counties with mandatory evacuations for the model, we considered the same as Laura with three additional counties to reflect the updated forecast track
(corresponding to 1,170,697 residents). The evacuation rates module now yields 1,054,500 expected evacuations [90\% CI: 832,500, 1,162,000] and 6,850 exportations [90\% CI: 4,100, 13,670]. Next, we compare the model's predictions with reported evacuations for Rita. According to a media survey \cite{lindner}, 2.5 million people evacuated, 59.9\% of which were \textit{not} residents in an evacuation risk zone. Therefore, approximately 1 million came from risk zones. Adjusting by population growth, this value is 15\% above our estimate. We also compare destinations. Of those respondents who evacuated, 32\% headed for Austin, San Antonio or Dallas/Fort Worth. Our estimated receptions for those cities is 323,000, which represents 30\% of the total receptions. Overall, these results show that the model's predictions are close to reported data, adding to the evidence that the model is a useful prediction tool for evaluating alternative scenarios.

\section{Discussion}
This study presents a framework that allows anticipating the impact of hurricane evacuations in an ongoing epidemic. The modules of the framework focus on predicting the size and dispersion of moving populations due to an approaching hurricane. Paired with prevalence numbers, these predictions produce estimates of disease exportation and importation numbers.

We underscore the framework's ability to capture spatial heterogeneity and incorporate geographic components that influence evacuation processes. Indeed, surge risk zones and preferred destinations are not homogeneously distributed geographically. Therefore, in the evacuation rates module, zones with higher surge risk are associated with higher participation rates. Similarly, in the destinations module, distance to highways and concentration of hotels are attributes that affect the preference for a certain destination in addition to distance between origin and destination counties. Thus, even if rigorous validation of the models' estimates remain to be evaluated, the point validations we present provide assurance that the models that integrate the framework generate likely evacuation patterns relative to the path of an approaching storm.

This study suggests few new directions for future research. First, an important limitation in the current framework is the omission of shadow evacuations in the evacuations module, which are those not coming from evacuation risk areas. Including them presents several challenges. For instance, the magnitude of the shadow evacuation for Hurricane Rita exceeded expectations  \cite{lindner}. There is general agreement that the exceptional response was influenced by the severity of Hurricane Katrina which preceded Rita by less than a month and had widespread media coverage  \cite{lindner}. Therefore, behavior change influenced by past recent events might dramatically affect evacuation predictions. Second, answering the question of whether the presence of an ongoing epidemic modifies evacuation decisions is out of the scope of this work. Nonetheless, it might be the case that instead of going to friends and family, evacuees decide to go to hotels to avoid spreading a disease. Finally, exportation and importation cases alone do not explain the transmission impact of evacuations. Future work should be concerned with the magnitude and variability of transmission increase due to importations and the overall temporal relocation of evacuees. 

Our framework’s flexibility and applicability is useful for disaster preparedness in the co-occurrence of hurricanes and epidemics. Combining the geographic characteristics' influence over the size of an evacuation with the specific characteristics of an epidemic can help authorities and emergency officials understand the transmission risk that may accompany a hurricane and guide emergency planning accordingly.

\begin{acknowledgement}
We thank Dr. Kelly Gaither and Dr. Gordon Wells of the University of Texas at Austin, as well as Mario Chapa from the Texas Division of Emergency Management, whose insights helped us better design and inform our models. 
\end{acknowledgement}
\section*{Appendix}
\section{Statistical Model for Predicting Evacuation Rates}\label{appendix:beta}

This section presents a simple weighted generalized linear model for predicting evacuation rates given a combination of risk zones and Hurricane categories.

We first introduce some notation. Let us denote $y_i$ the registered evacuation rate for all $i=1,\hdots,N$. Similarly, let $z_i\in\{0,\hdots,5\}$ be the indicator of risk zone for and $h_i\in\{0,\hdots,5\}$ the indicator of hurricane intensity.

Then, the statistical model is described as the following generalized Beta regression parameterized by a mean and precision parameter
\begin{equation*}\label{eq:evac-rate}
\begin{aligned}
y_i &\sim \mathrm{Beta}(\mu_i \lambda, (1 - \mu_i) \lambda) \\
\textrm{logit}(\mu_i) & = \alpha + \beta^\text{zone}_{z_i} + \beta^\text{intensity}_{h_i}.
\end{aligned}
\end{equation*}
where $\theta:=(\alpha, \{\beta^\text{zone}_j\}_{j=0}^5, \{\beta^\text{intensity}_k\}_{k=0}^5)$ are learnable parameters. Using this parameterization
$
\mathbb{E}[y_i]=\mu_i$
and $\mathrm{Var}[y_i] = \mu_i(1 - \mu_i) / (\lambda + 1)$.

Finally, we note that observations carry different weights depending on whether or not the evacuation rate is observed or intended (see Section \ref{subsec:evac_module}). For this purpose, we let $\omega_i$ be a confidence weight such that $\omega_i=1$ if $y_i$ is the report from actual observed evacuation rate, and $\omega_i=1/2$ if the data source is a reported intention to evacuate. Then $\theta$ is obtained by a weighted maximum likelihood with the weights defined by $\{\omega_i\}_{i=1}^N$.

\section{Multinomial Logit Model for Origin-destination Prediction}\label{appendix:od}

Let use denote by $P_{ij}$ the probability of choosing destination $i$ given origin $j$. The model takes the following form:
\begin{equation*}
    P_{ij} = \frac{\exp(\beta^\top x_{ij})} {\sum_{k=1}^K\exp(\beta^\top x_{ik})}
\end{equation*}
where the $x_{ik}$'s are the vectors of geographic, demographic and infrastructure features representing destination $k$ and the cost of traveling from $i$ to $k$ for $k=1,\hdots,K$ alternatives. These features are listed in Table \ref{tab:logit}.

\end{document}